\documentclass[apl,reprint]{revtex4-1}
\usepackage{chemformula}
\usepackage{graphicx}
\usepackage{siunitx}
\usepackage{float}

\begin{document}

\title{From 2D to 3D: graphene moulding for transparent and flexible probes}

\author{Martin Lee}
\affiliation{Department of Physics, McGill University, Montreal, H3A 2T8, Canada}
\author{Thai-An Vuong}
\affiliation{Department of Physics, McGill University, Montreal, H3A 2T8, Canada}
\author{Eric Whiteway}
\affiliation{Department of Physics, McGill University, Montreal, H3A 2T8, Canada}
\author{Xavier Capaldi}
\affiliation{Department of Physics, McGill University, Montreal, H3A 2T8, Canada}
\author{Yuning Zhang}
\affiliation{Department of Physics, McGill University, Montreal, H3A 2T8, Canada}
\author{Walter Reisner}
\affiliation{Department of Physics, McGill University, Montreal, H3A 2T8, Canada}
\author{Carlos Ruiz-Vargas}
\affiliation{Department of Physics, McGill University, Montreal, H3A 2T8, Canada}
\author{Michael Hilke}
\email[corresponding: ]{hilke@physics.mcgill.ca}
\affiliation{Department of Physics, McGill University, Montreal, H3A 2T8, Canada}


\begin{abstract}

Chemical vapor deposition (CVD) has been widely adopted as the most scalable method to obtain single layer graphene. Incorporating CVD graphene in planar devices can be performed via well established wet transfer methods or thermal adhesive release. Nevertheless, for applications involving 3D shapes, methods adopted for planar surface provide only a crude solution if a continuous, tear-free, wrinkle-free graphene layer is required. In this work, we present the fabrication and characterization of PDMS-supported 3D graphene probes. To accommodate 3D geometries, we perform CVD on catalysts possessing a non-trivial 3D topology, serving to mold the grown graphene to a final non-trivial 3D shape. This advance overcomes challenges observed in standard transfer processes that can result in uneven coverage, wrinkles and tears. To demonstrate the potential of our new transfer approach, we apply it to fabricate graphene electrical probes. Graphene, due to its flexibility, transparency and conductivity, is an ideal material with which to replace conventional metal based probes. In particular, with a contact resistance in the order of tens of k$\Omega$, these graphene probes may find applications, such as in electrophysiology studies.  

\end{abstract}

\maketitle

\section{Introduction}

Graphene is a two dimensional carbon allotrope widely known for its remarkable and versatile properties: it is flexible \cite{stoberl2008morphology}, stretchable \cite{kim2009large}, conductive \cite{neto2009electronic}, transparent \cite{nair2008fine}, yet extremely strong \cite{lee2008measurement}. Graphene is consequentially an ideal material for flexible electronics \cite{cooper2012experimental}. However, current manufacturing processes are limited to producing flat graphene samples as the catalyst that graphene is grown is typically obtained in the form of planar substrates \cite{li2009large}. Furthermore, the transfer steps necessary to move graphene from the catalyst surface to the final device complicate the fabrication process and introduce defects from the polymer support layers \cite{pirkle2011effect}. The most widely used graphene transfer process \cite{li2009transfer,jiao2008creation,reina2008transferring} entails coating a layer of poly(methyl-methacrylate) (PMMA) on top of graphene, etching away the catalyst to create a graphene-PMMA stack, transferring the stack to a desired substrate and dissolving the PMMA with acetone to leave only the graphene. While this process is simple and efficient, it introduces cracks, wrinkles, contamination and defects in the transferred graphene \cite{wang2016direct}, significantly compromising final device quality.

Due to graphene's unique properties, many processes have been developed to incorporate graphene, a two dimensional material, in three dimensional structures. In particular, novel 3D structures made from graphene foam exhibit higher electrical conductivity than planar graphene-based composite does and demonstrate a great potential for flexible and stretchable conductors \cite{chen2011three}. Composites of 3D graphene foam with inorganic compound are widely studied and have high reversible capacity, outstanding cycling rate capability in lithium batteries \cite{wei20133d} and demonstrate remarkable performance as an enzymeless glucose sensor \cite{dong20123d}. However, some fabrication strategies for 3D graphene involve extremely tedious fabrication process while being exclusive only to a specific purpose. Furthermore, those methods are limited to produce small scale devices and inevitably hinder a wide range of applications. Due to this lack of versatility, these approaches are not suitable for industrial scaling and are consequently limited to research use.

Here we present an industrially scalable approach for fabricating three dimensional graphene structures. We utilize a catalyst, copper, purchased in the form of a cylindrical tube (so that is already moulded into the desired final shape), and perform CVD to grow graphene on the tube surface. We then fill the inside of the tube with a curable polymer for structural support and etch away the copper. Using this method, we successfully fabricated graphene based flexible probes. Furthermore, the geometry facilitates the addition of fibres directly into the probe, providing optical access for combined contact electro-optical applications. Distinct from other approaches where 3D graphene is grown in a foam-like structure, the graphene incorporated in our device is layered graphene, conforming to the device's three dimensional shape. The shape of the copper mould used for final device is completely flexible. As such, the final device is fully modifiable, providing a wide array of applications. The curable polymer used for structural support can be chosen freely on an application specific basis.

\section{Fabrication}

\begin{figure*}[!ht]
\includegraphics[width=0.9\textwidth]{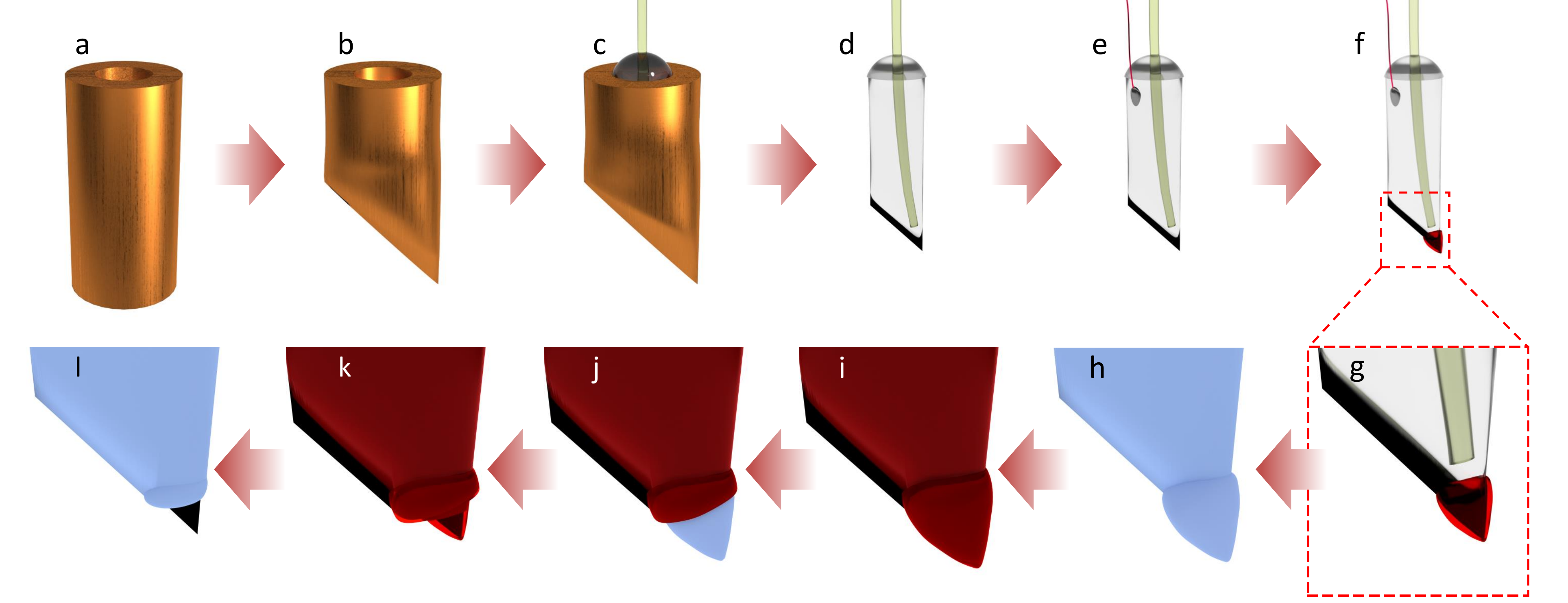}
\caption{Illustration of the graphene probe fabrication steps. (a) The starting substrate is a copper capillary tube. (b) One end of the tube is snipped at an angle and graphene is grown on the tube's surface. (c) The tube inside is coated with PMMA, baked and filled with PDMS. After dessication, an optical fibre is inserted. (d) The copper is etched away in ammonium persulfate. (e) A contacted wire is attached to the graphene via silver epoxy. (f) The tip of the probe is dipped in AZ 9245 using an xyz stage. (g) Zoomed in image of (f). (h) Parylene C is deposited isotropically. (i) The entire probe is coated with S1805. (j) The tip is exposed and developed. (k) O$_2$ plasma is used to remove the parylene. (l) The photoresist is removed by acetone.}
\label{fig:fab}
\end{figure*}

The fabrication protocol is illustrated in figure \ref{fig:fab}. 
Copper capillary tubes with an inner diameter of \SI{600}{\micro \meter} and outer diameter of \SI{800}{\micro \meter} are cut into \SI{2.5}{\centi \meter} sections (figure \ref{fig:fab} (a)) using a crimper. The copper tubes are first sonicated in RCA SC-1 solution (1 NH$_4$OH:1 H$_2$O$_2$:5 DI) for 30 minutes. Next, the tubes are rinsed in deionized water (DI) and sonicated in 4.2 \% HNO$_3$ for 30 mininutes and rinsed in DI once more. One end of the tubes are then cut at an angle ($\leq$ 45 $^\circ$ with respect to the body) which creates a sharp tip (figure \ref{fig:fab}(b)) and rinsed in 2-propanol. The copper tubes are blow dried and placed in our home-made vertically loaded CVD system. The tubes are annealed at 1000$^\circ$C for $\geq$2 hours while flowing 20 sccm of H$_2$ gas. The temperature is then increased to 1050$^\circ$C and 4 sccm of CH$_4$ are introduced. After 1 hour of growth the tubes are removed from the furnace to cool down while maintaining the gas flow.

Once the graphene coated tubes are removed from the CVD, they are submerged in PMMA A2 in a desiccator to promote a complete coating. We observed that the intermediate PMMA layer between graphene and PDMS is necessary because the PMMA A2 is less viscous, and therefore conforms better to the surface morphology of the graphene covered copper than PDMS alone. The tubes are then individually spun at 1000 rpm with the open-end pointing out to remove the excess PMMA and ensure a uniform coating $\sim$ \SI{100}{\nano\meter}. The probes are then baked at 120$^\circ$C for 2 minutes and cooled down. PDMS (Sylgard 184) is mixed (1 curing agent:10 base) and dessicated for 45 minutes. Once all the bubbles escape, the graphene coated tubes are submerged in the PDMS mix and further dessicated to ensure a void-free structure. After the PDMS settles in the tubes, \SI{100}{\micro \meter} optical fibers (Edmund Optics 57-061) are inserted in the copper mould as shown in figure \ref{fig:fab} (c). The tubes are then baked at 80$^\circ$C for 1 hour to cure the PDMS.

Before the copper can be etched, PDMS on the outside of the copper tube is scraped off. Using the optical fiber as support, the tubes are then immersed in a bath of 5\% w.t.(NH$_4$)$_2$S$_2$O$_8$ heated to 80$^\circ$C. Once the copper has been fully etched (figure \ref{fig:fab} (d)), the probes are rinsed in DI three times and blown dry. As shown in figure \ref{fig:fab} (e), small amounts of conductive silver epoxy (H2OE EPO-TEK Ted Pella) are used to contact a wire onto the side of the graphene probe body for electrical measurement. 

We also provide a possible method for patterning the graphene probe with standard lithography techniques shown in figure \ref{fig:fab}(f) to (l). Prior to doing so, we characterized the probe using SEM and Raman spectroscopy and measured the probe electrical impedance.

\begin{figure}[ht]
\centering
\includegraphics[width=0.9\columnwidth]{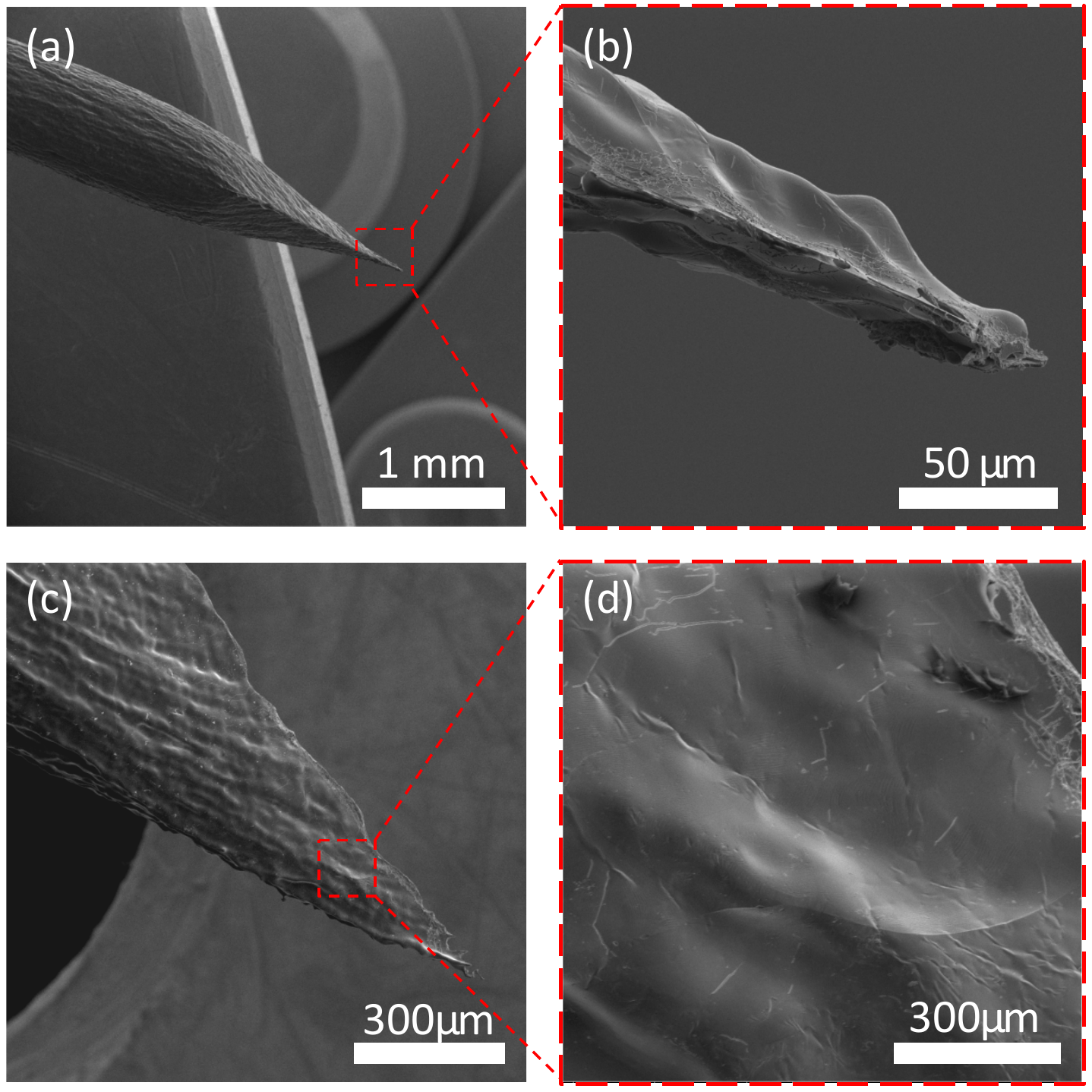}
\caption{SEM images of the graphene probe at various magnifications: (a) viewed from the cut edge of the graphene probe and (b) same at high magnification; (c) viewed from the side of the graphene probe and (d) same at high magnification. }
\label{fig:sem}
\end{figure}
The SEM images of the three dimensional structures completely covered by graphene - equivalent to the fabrication step in \ref{fig:fab}(d) - is displayed in figure \ref{fig:sem}. Figure \ref{fig:sem} (a) and (c) are graphene probes rotated 90 degrees from another, (a) showing the graphene probe seen from the cut edge and (c) showing the graphene probe from the side. Respective images in high magnification are shown in (b) and (d). These images show that the graphene probe closely follows the surface contour of the copper mould. The tip of the graphene probe, as seen in figure \ref{fig:sem} (a), (b) and (c) converges to a point as sharp as a few tens of microns in curvature.

These SEM images show wrinkles and surface roughness of the graphene arising from the copper substrate. This alludes to the retention of the shape of the probe even after the copper is etched away aided by the underlying polymer. Furthermore, a minimal number of rips and tears are visible under SEM. The tears that are present can be attributed to the lack of a transfer step from the copper to the target substrate.

We further developed a method to insulate the graphene probe while only exposing the tip (figure \ref{fig:fab}(f) to (l)). A drop of AZ-9245 on a glass slide is placed on a xyz stage and the probe is hung from above by the optical fiber. The xyz stage is slowly raised until the probe and AZ-9245 are in contact with a meniscus formed at the interface (figure \ref{fig:fab} (f), (g)). The probe is left to dry for an hour and suspended in the SCS200 parylene coater for a uniform isotropic deposition. One gram of parylene C dimers are used to deposit 580 nm parylene on the probe surface (figure \ref{fig:fab} (h)). After deposition, the probe is dipped in S1805 and suspended to dry for an hour (figure \ref{fig:fab} (i)). Aluminum foil is used to mask the probe body so that only $\sim$\SI{50}{\micro\meter} region around the tip is exposed. An EVG-620 aligner is used to expose the S1805 coated probe (450 J/cm$^2$). The probe is then developed in MF-319 for 100 seconds, revealing the parylene at the tip (figure \ref{fig:fab} (j)). Using an O$_2$ plasma asher, the parylene at the tip is removed (figure \ref{fig:fab} (k)). Finally, the probe is submerged in acetone to remove the photoresist masking layer (figure \ref{fig:fab} (l)). 

\begin{figure}[ht!]
\centering
\includegraphics[width=0.8\columnwidth]{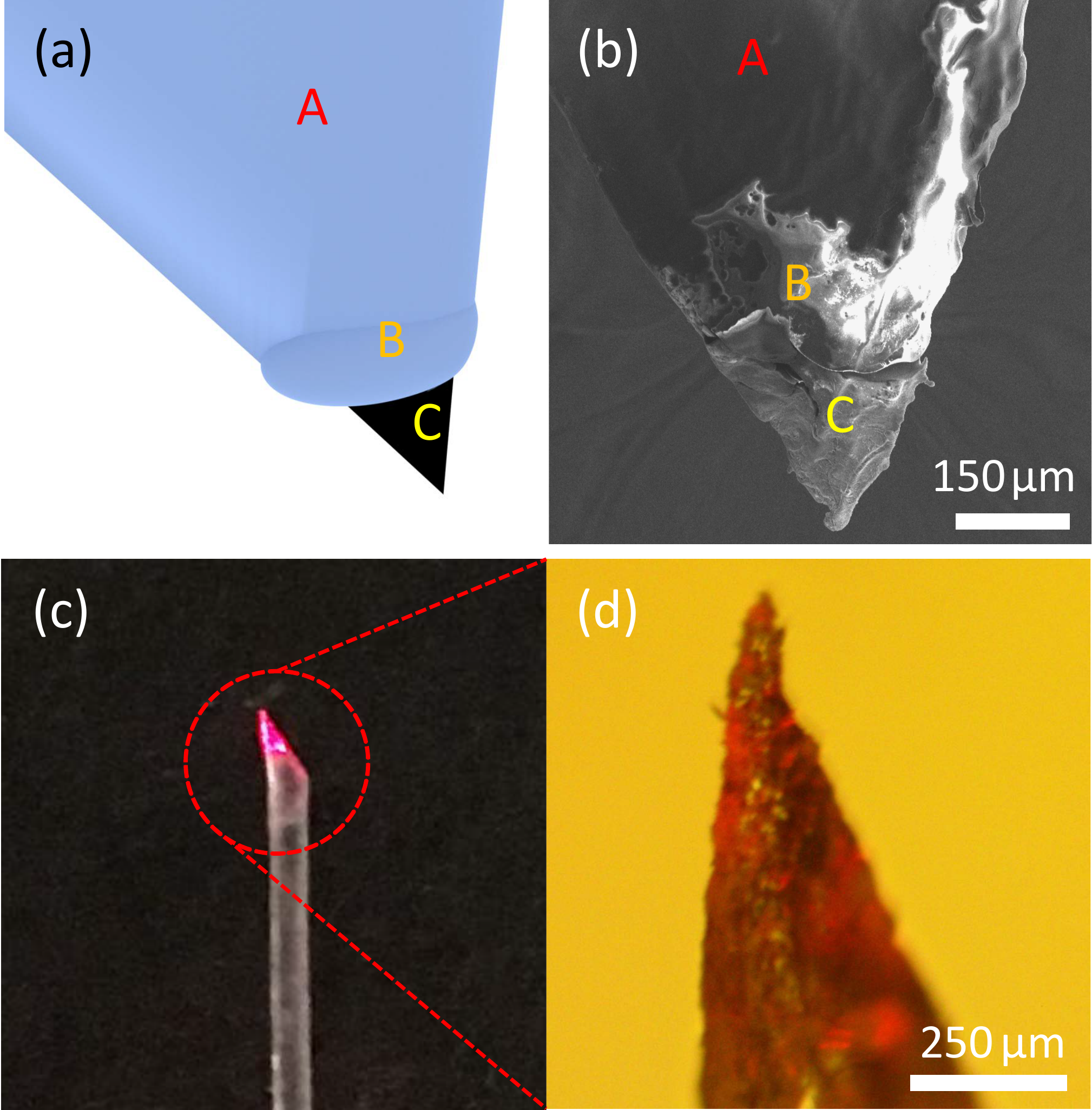}
\caption{(a) Illustration of the probe with exposed tip compared to (b) the actual device. A: Parylene on graphene. B: Parylene on AZ 9245 on graphene. C: Graphene. (c) Optical image of the probe while light with a wavelength of 650 nm is shone through the optical fiber. A 10 mW visual fault locator is used as a light source.}
\label{fig:discussion}
\end{figure}

An SEM image of an exposed tip is shown in figure \ref{fig:discussion} (b) compared to an illustration in (a). The exposed graphene region is approximately \SI{200}{\micro\meter} with no photoresist residue on its surface. All regions are separated with clear boundaries, suggesting a high precision in the proposed patterning method.

\section{Characterization}

\begin{figure}[ht]
\includegraphics[width=\columnwidth]{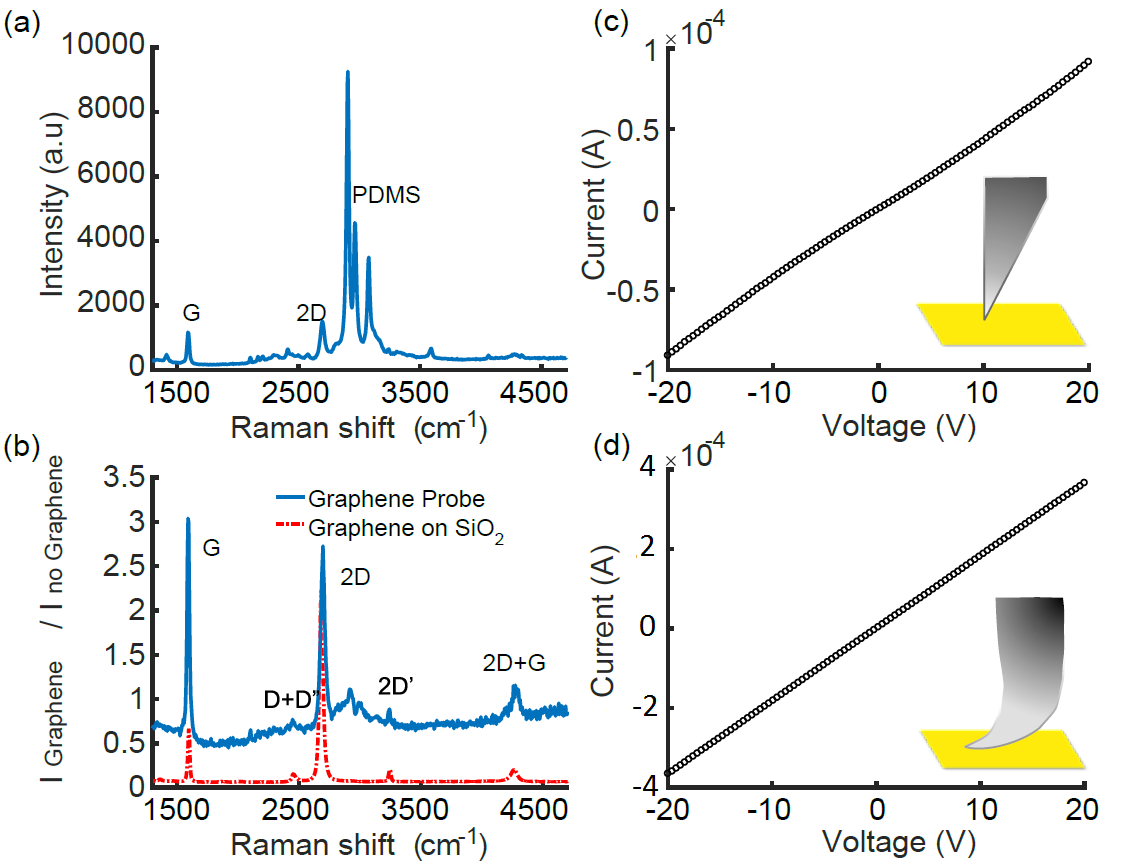}
\caption{(a-b) Raman spectroscopy performed on the body of the graphene probe and I-V curves. (a) Raman spectrum from the body of the graphene probe before the parylene deposition. (b) A region containing graphene and a region without graphene plotted in a ratio (blue) compared to averaged Raman spectrum of graphene on SiO$_2$ (red). (c) IV trace of a graphene probe in point contact with a gold pad. (d) IV trace of a graphene probe in areal contact with a gold pad.}
\label{fig:iv}
\end{figure}

To reveal the presence and quality of graphene on the tips we performed Raman spectroscopy. 
A Renishaw InVia Raman system with wavelength of 514 nm was used on the body of the graphene probe. As shown in figure \ref{fig:iv}(a), the Raman spectrum is heavily dominated by the PDMS peaks. However, the PDMS peaks are not in superposition with the G peak and the 2D peak of graphene thus allowing for post-fabrication characterization of graphene. The ratio of the 2D peak and the G peak is $\sim$0.86, which suggests that many domains of multilayer graphene are present \cite{graf2007spatially,yu2009large}. This is further supported by the SEM images in figure \ref{fig:sem}(c) and (d). We then removed the PDMS contribution by subtracting the spectrum taken for an area of the probe not containing graphene. This revealed various peaks of graphene such as the D+D" peak, 2D' peak and 2D+G peak \cite{bernard2012probing}. This spectrum was compared to the spectrum from a typical two dimensional graphene on SiO$_2$/Si as shown in \ref{fig:iv}(b).

Next, we performed electrical measurements on the graphene probes using a Keithley 2400 sourcemeter. The I-V curves of a graphene probe device are shown in figure \ref{fig:iv}. Probes are brought into contact with a gold pad evaporated on glass and the voltage is swept while the current is measured. As shown in figure \ref{fig:iv} (c), the probes have a slight non-linear I-V behaviour when only the tip is in contact with the gold pad at low tip-substrate pressure. When greater force is used the probe tip bends and conforms to the surface of the gold pad resulting in a larger contact area. Resulting linear I-V trace is shown in figure \ref{fig:iv} (d). These results are based on the fabrication of about 100 tips with a yield of close to 100\%. The resistance ranges between 15 and 250 k$\Omega$ when the tip barely touches the gold (point contact), while the resistance reaches approximately 10-50 k$\Omega$ when the tip is pushed into the gold surface under pressure (areal contact).

\section{Discussion}

Graphene is grown on the surface of commercially available polycrystalline low purity copper capillary tube. Low purity copper will induce a large number of nucleation sites for the graphene CVD growth \cite{braeuninger2016understanding}. This leads to faster growth of graphene and also a prevalence of multilayer regions. For most applications this is not an issue, since the conductivity is not strongly affected (doping plays a more dominant role). However, if larger graphene crystals domain sizes are desired higher purity copper molds can be used. This is relevant for high mobility electronic devices such as Hall sensors, which require high mobilities (the mobility scales close to linearly with average grain size \cite{van2013scaling}). The effect of purity on the mechanical properties are not as well understood. Having multiple multilayer regions can also enhance some mechanical properties \cite{Choi:2010aa}.

The mould geometry has the highest impact on our probe quality, as the copper substrate entirely determines the geometry of the resulting device. We formed the sharp probe tip by closing off one side of the capillary copper tube. However, designing a copper substrate with a desired three dimensional shape would provide full control over the probe shape. With the continuously increasing affordability of 3D printers, we believe graphene moulding can easily be implemented in many complex applications, including but not limited to medicine and artificial tissues \cite{podila2013graphene}, where graphenes biocompatibility and hemocompatibility properties are needed \cite{pinto2013graphene}. Traditional approaches such as metal injection could also be used to create the copper mold \cite{moballegh2005copper}. The copper quality will determine the graphene growth parameters (such as growth time and graphene characteristic) and the copper volume the etching time required.

One potential application of our probe is electrophysiology, where rigid metal rods are used to probe the electrical activities of nervous tissues. However, these rigid rods are not able to conform to the tissue of interest, causing swelling of the nearby glial cells or damage to the neurons \cite{grill2005safety}. The flexibility of the graphene probe allows for a conformal contact onto the target surface leading to a lower impedance while reducing the possible damage to nearby tissues. In order to minimize the leakage of current through the ionic solution - which decreases the signal to noise ratio - we provided a possible method of patterning the tip of the graphene probe in figure \ref{fig:fab} (e) to (l) with the finished product shown in figure \ref{fig:discussion} (a) and (b). 

Furthermore, with the introduction of the optical fibre, our probes can be implemented in optogenetics studies \cite{aravanis2007optical}, delivering optical access simultaneously with electrical access at near single cellular level. Figure \ref{fig:discussion} (c) and (d) show optical images of the probe connected to a 650 nm light source. The optical fibre successfully delivers light to the tip of the probe.

\section{Conclusion}
We demonstrated the fabrication and characterization of few layer graphene flexible electro-optical probes using CVD and copper based moulding. This process can be generalized to other shapes and high throughput. The graphene layers are largely defect free and are composed of mostly single layers with some regions with few layers as probed by Raman spectroscopy and scanning electron microscopy.  The electrical properties were characterized using I-V measurements, which show close to linear I-Vs. The resistance is stable between measurement cycles and varies with contact pressure due to the increased contact area. The resistance is of the order of a few tens of kilo-ohms. Possible applications of these probes include in-vivo electrophysiology, optogenetics and optical fiber based Raman spectroscopy of nervous tissues using the flexible graphene probe \cite{grill2005safety}.


\bibliography{refs}

\end{document}